# Giant nonlinear anomalous Hall effect induced by spin-dependent band structure evolution

Xiangyu Cao[1†], Jie-Xiang Yu[2,3†], Pengliang Leng[1†], Changjiang Yi[4,5†], Yunkun Yang[1], Shanshan Liu[1], Lingyao Kong[6], Zihan Li[1], Xiang Dong[1], Youguo Shi[4], Jiadong Zang[2*], Faxian Xiu[1,7,8,9*]

[1] State Key Laboratory of Surface Physics and Department of Physics, Fudan University, Shanghai 200433, China

[2] Department of Physics and Astronomy, University of New Hampshire, Durham, New Hampshire 03824, USA.

[3] Department of Physics, University of Florida, Gainesville, Florida 32611, USA

[4] Beijing National Laboratory for Condensed Matter Physics, Institute of Physics, Chinese Academy of Sciences, Beijing 100190, China

[5] University of Chinese Academy of Sciences, Beijing 100049, China

[6] School of Physics and Materials Science, Anhui University, Hefei, 230601, PR China

[7] Institute for Nanoelectronic Devices and Quantum Computing, Fudan University, Shanghai 200433, China

[8] Collaborative Innovation Center of Advanced Microstructures, Nanjing 210093, China

[9] Shanghai Research Center for Quantum Sciences, Shanghai 201315, China

† These authors contributed equally to this work
* Correspondence and requests for materials should be addressed to F. X. (E-mail: Faxian@fudan.edu.cn) or J. Z. (E-mail: Jiadong.Zang@unh.edu)

**Abstract:**

**Anomalous Hall effect (AHE) is the key transport signature unlocking topological properties of magnetic materials. While AHE is usually proportional to the magnetization, the nonlinearity suggests the existence of complex magnetic and electron orders. Nonlinear AHE includes the topological Hall effect (THE) that has been widely used to identify the presence of spin chirality in real space. But it can in principle be induced by band structure evolution via Berry curvatures in the reciprocal space. This effect has been largely overlooked due to the intertwined mechanisms in both real and reciprocal spaces. Here, we observed a giant nonlinear AHE with the resistivity up to 383.5 $\mu\Omega \cdot cm$, contributing unprecedentedly 97% of the total Hall response in $EuCd_2As_2$. Moreover, it can be further enhanced by tilting the magnetic field 30° away from [001] direction, reaching a large anomalous Hall angle up to 21%. Although it shows exactly the same double-peak feature as THE, our scaling analysis and first-principles calculations reveal that the Berry phase is extremely sensitive to the spin canting, and nonlinear AHE is a consequence of band structure evolution under the external magnetic fields. When the spins gradually tilt from the in-plane antiferromagnetic ground state to out-of-plane direction, band crossing and band inversion occur, introducing a bandgap at $\Gamma$ point at a canting angle of 45°. That contributes to the enhancement of Berry curvature and consequently a large intrinsic Hall conductivity. Our results unequivocally reveal the sensitive dependence of band structures on spin tilting process under external magnetic fields and its pronounced influence on the transport properties, which also need to be taken into consideration in other magnetic materials.**

# I. INTRODUCTION

The anomalous Hall effect (AHE) is vital transport signature connecting electron properties and magnetic orders in magnetic materials. While it is assumed to show a linear relation with magnetization (M) in most cases [1], the nonlinearity of AHE to M is a signature of novel magnetic and/or electron orders [1,2]. The most celebrated nonlinear AHE is the topological Hall effect (THE), which can be viewed as a special kind of AHE induced by real-space spin chirality [2–8]. In non-coplanar spin textures such as magnetic skyrmions, the non-zero scalar chirality $\chi_{ijk} = S_i \cdot (S_j \times S_k)$ generates a finite real-space Berry phase, acting as an effective magnetic field on conduction electrons [3], and leads to the topological Hall resistivity. Due to the unique pronounced peak feature, it has been widely used as an exclusive fingerprint of chiral spin structures [9,10]. On the other hand, the intrinsic part of AHE is induced by momentum space Berry curvature, which can in principle deviate from linearity as reported in a limited number of systems [11–17]. Nonlinear AHE can thus take place once the band structure is sensitive to the magnetization evolution. But since these two mechanisms both exhibit two opposite peaks during the magnetization, it is extremely difficult to distinguish one from the other [18–20]. Furthermore, complex spin structures could strongly affect band structures, so these two mechanisms are very often intertwined and a clear separation is hardly achievable.

Here we report a giant nonlinear AHE showing pronounced peaks in EuCd$_2$As$_2$ with the largest resistivity $\rho_{xy}^{NA}$=383.5 μΩ·cm, which takes up to 97% of the total Hall resistivity and fully dominates the Hall response. It shows exactly the same feature as THE in chiral spin textures. But surprisingly, our analysis shows a momentum-space Berry curvature origin, as further supported by the detailed scaling analysis between longitudinal and Hall resistivities, and first-principles calculations of band structure evolution process. The spin orientations play an important role in the evolution of band structure, causing band inversions and gap opening at certain canting angles, which results in significant enhancement of Berry curvature. Besides, the $\rho_{xy}^{NA}$ and the anomalous Hall angle can be further enhanced by tilting the magnetic field 30° away from the [001] direction. Particularly at low temperature T=2K, both values are elevated from 100 μΩ·cm and 5% to 430 μΩ·cm and 21%, respectively, indicating that the momentum Berry curvature changes with not only the intensity but also the direction of the external magnetic field. Our results unequivocally demonstrate the typical THE feature cannot be directly linked to the chiral spin structures, and the special AHE response in Weyl semimetals is not always originated from the Weyl nodes. More than that, we clearly reveal how spin textures influence the band structures, resulting in an unexpected sensitive dependence of momentum-space Berry phase on spin orientations, whose influence on not only AHE but also other transport properties cannot be neglected in magnetic materials.

# II. RESULTS

EuCd$_2$As$_2$ was theoretically predicted and experimentally demonstrated to be a magnetic Weyl semimetal [21–24]. As shown in Fig. 1(a), EuCd$_2$As$_2$ has a trigonal crystal structure with a space group P$\bar{3}$m1 (no. 164). The crystal has a layered structure consisting of alternative triangular Eu layers and Cd$_2$As$_2$ bilayers. Eu atoms contribute to the magnetism in this system, forming an in-plane A-type antiferromagnetic (AFM) structure below the Néel temperature T$_N$=9.5K [21,25]. We performed first-principles calculations to confirm the magnetic properties. According to the total energy analysis, the antiferromagnetic spin-ordering is 0.8 meV (9.3 K) lower in total energy than ferromagnetic spin ordering. The spin-ordering with the in-plane Néel vector is 0.39 meV in total energy lower than the out-of-plane Néel vector. The in-plane antiferromagnetic spin-ordering is thus the ground state, consistent with previous experimental observations [21].

As the central result of this work, the detailed temperature-dependent transport measurements with the external magnetic field applied along [001] direction are shown in Figure 1(b-f). The Hall resistivity $\rho_{xy}(B)$ in Fig. 1(b) shows the prominent double-peak feature. With the temperature increasing from 2K to 50K, the peaks are first enhanced, reaching a maximum value at T=9.5K, then suppressed at higher temperatures and almost invisible at T>50K. Compared to the M(B) curve in the Supplementary Materials, the pronounced two peaks near zero magnetic field in the original $\rho_{xy}(B)$ curves clearly deviate from linearity in magnetization. But it is almost impossible to confirm its origin only with the experiment data, so the total Hall resistivity is expressed as $\rho_{xy} = R_0 B + R_s M + \rho_{xy}^{NA}$, where the $\rho_{xy}^{NA}$ term represents the AHE nonlinear to M. $\rho_{xy}^{NA}$ with two peaks observed here is so typical to be conventionally regarded as the footprint of typical THE, but it can contain intrinsic AHE from momentum-space Berry phase as well. The original Hall signal at T=9.5K [Fig. 1(c)] shows a giant nonlinear AHE (red zone), which can be obtained after subtracting the ordinary Hall effect (OHE) and the conventional AHE (blue line) from the original Hall resistivity (red line) by fitting the original Hall data from B=4T to 6T to the equation $R_0 B + R_s M$. The maximum $\rho_{xy}^{NA} = 383.5\ \mu\Omega\cdot\text{cm}$ appearing at B*=±0.2T is 5 orders of magnitudes larger than the saturation anomalous Hall resistivity. Furthermore, the nonlinear AHE ratio defined as $\frac{\rho_{xy}^{NA}}{|\rho_{xy}^O|+|\rho_{xy}^A|+|\rho_{xy}^{NA}|}$ (equal to $\rho_{xy}^{NA}/\rho_{xy}$ for our system) reaches 97%, indicating that this feature completely dominates the Hall response at the peak position. The $\rho_{xy}^{NA}$ at different temperatures [Fig. 1(d)] were extracted by the same procedure. Figure 1(e) presents the nonlinear AHE ratio and the calculated anomalous Hall angle $\theta_{AH} = 1 - \sigma_{xy}^O/\sigma_{xx}$ at different temperatures. It always contributes more than 75% of the total Hall signal below 15K and the maximum Hall angle reaches 15%, showing a prevailing response. In Fig.1(f), we compare our results with prior reports of nonlinear AHE systems, including THE systems [4–8,26–33] and nonlinear intrinsic AHE [14,15]. The data of the related systems are estimated from refs. [4-8,14,15,26-33]. The diagram explicitly shows the salient nonlinear AHE feature in EuCd2As2 with a giant contribution ratio and $\rho_{xy}^{NA}$ (except the insulating systems [34,35]).

The nonlinear AHE in EuCd2As2 shows exactly the same feature with THE originated from the real-space Berry phase and can be conventionally regarded as important evidence of chiral spin texture. Eu atoms have a magnetic moment closed to $7\mu_B$ [21], showing a large exchange interaction to conduction electrons, so the adiabatic limit is respected. If net chirality were to exist, it must contribute to THE. However, EuCd2As2 has a point group $\bar{3}m1$ so that the spatial inversion is a lattice symmetry. Furthermore, the midpoint between any pair of neighboring Eu atoms is always an inversion center, so that the Dzyaloshinskii-Moriya interaction is absent everywhere. The only possible spin canting takes place at the domain wall. Prior X-ray observation indicates the possible presence of magnetocrystalline anisotropy in the plane, leading to three domains pointing 120 degrees away from each other [21]. Once an external magnetic field is applied, spins tilt up, and spin canting emerges at the triple point where three domains meet together. However, since three domains have equal populations, we have the same numbers of triple points with in-plane spins rotating clockwise and counter-clockwise, so the overall scalar chirality is zero. Actually, moving along a domain wall from any triple point, one always arrives at another triple point where spin rotates oppositely [Fig. 2(a)]. The Monte Carlo calculation supports this argument. As shown in Fig. 2(b), at finite temperatures, populations of topological charges $\pm Q$ are the same. We thus conclude the absence of the real-space Berry phase in this system, and this nonlinear AHE is impossible to be the real THE.

This conclusion is further supported by the scaling analysis of transport measurement results. For the real-space-Berry-phase driven THE, the resistivity is related to the scalar chirality generated emergent magnetic field $B_{eff}$ by $\rho_{xy}^T = \frac{B_{eff}}{ne}$. So it does not have a scaling relation to the longitudinal resistivity $\rho_{xx}$. On the other hand, the momentum-space Berry phase contributes to intrinsic Hall conductivity $\sigma_{xy}^I = \frac{e^2}{\hbar}\sum_n \int \frac{d\mathbf{k}}{(2\pi)^3} b_n(\mathbf{k})$, where $b_n(\mathbf{k})$ is the Berry curvature of band $n$. It contains both the anomalous Hall conductivity $\sigma_{xy}^A$ linear in magnetization M and nonlinear AHE related to the band structure. It contributes to the Hall resistivity by $\rho_{xy}^I = \sigma_{xy}^I/(\sigma_{xy}^2 + \sigma_{xx}^2)$. In metallic systems [1], $\sigma_{xx} \gg \sigma_{xy}$, so that as long as the band structure does not change, the Hall resistivity $\rho_{xy}^I$ is proportional to $\sigma_{xx}^{-2} \sim \rho_{xx}^2$. In EuCd$_2$As$_2$ with dominant $\rho_{xy}^{NA}$, $\rho_{xy}^{NA} \propto \rho_{xx}^2$ is expected if the momentum-space Berry phase prevails [1]. The temperature-dependent $\rho_{xy}^{NA}$ (from Fig. 1) as a function of $\rho_{xx}^2$ is shown in Fig. 2(c). The data measured at T<T$_N$ fit very well to the linear scaling, giving strong evidence of the momentum-space origin of nonlinear AHE in EuCd$_2$As$_2$. But for T>T$_N$, the magnetic ground state and the band structure are temperature dependent. So the data measured at T>T$_N$ gradually deviate from the linear relation. Noticeably, two peaks are observed in $\rho_{xx}(B)$ data, as shown in Fig. 2(d), possibly due to the enhanced scattering in the spin polarization process. To rule out the contribution from trivial $\rho_{xx}$ (or $\sigma_{xx}$) variance, we decompose the measured resistivity to $\sigma_{xx}$ and $\sigma_{xy}$ in Fig. 2(e). Both nonlinear anomalous Hall conductivity $\sigma_{xy}^{NA} = \rho_{xy}^{NA}/(\rho_{xx}^2 + \rho_{xy}^2)$ and the total non-ordinary Hall conductivity $\sigma_{xy}^{A+NA} = \sigma_{xy}^A + \sigma_{xy}^{NA}$ are presented. Note that $\sigma_{xy}^A = R_s M/(\rho_{xx}^2 + \rho_{xy}^2)$ here only represents AHE linear to M and contains both intrinsic and extrinsic contributions. $\sigma_{xy}^{NA}$ still shows a prominent peak, indicating that the nonlinear AHE in this system is an intrinsic behavior. The calculated $\sigma_{xy}^{NA}(B)$ at different temperatures are shown in Fig. 2(f). The peak values in $\sigma_{xy}^{NA}(B)$ are almost independent of temperature for T<T$_N$, which is another strong evidence since the band structure, and consequently the momentum-space Berry curvature, does not change much with temperature. It is in sharp contrast to $\rho_{xy}^{NA}$ and $\theta_{AH}$ whose dramatic temperature dependence in Fig. 1(d-e) mainly comes from the extrinsic $\sigma_{xx}$ due to the temperature-dependent spin scattering.

If the field-dependent $\sigma_{xy}^{NA}$ is the intrinsic Hall conductivity, the most likely source is the band structure variation induced by non-collinear spin cantings. This is confirmed by the first-principles calculations. A series of spin configurations with various canting angles have been prepared. Figure 3(a) shows the band structure and the calculated intrinsic Hall conductivities $\sigma_{xy}^I$ when the canting angle is $45°$. A prominent peak of $\sigma_{xy}^I$ has been found at $E = -84$ meV. This peak position should be closer to the real Fermi surface of the sample by the fact that our sample is heavily p-doped [24] with a large carrier density $n_p = 1.46 \times 10^{19} cm^{-3}$ at T=2K extracted from the ordinary Hall effect data. Peak of $\sigma_{xy}^I$ thus contributes significantly to the measured Hall conductivity. Hall conductivities in such a Weyl semimetal are usually attributed to the amplified Berry curvature around the Weyl nodes. But here the Weyl node above zero energy is far from the Fermi surface and should not affect the transport properties. Figure 3(b) shows the evolution of band structures along $K - \Gamma - A$ as the canting angle changes. Here only bands around the Fermi energy are marked in order to clearly reveal the band evolution process and avoid distraction from irrelevant band hybridizations. At the collinear antiferromagnetic ground state with zero canting angle, all bands are degenerate as protected by the Kramers degeneracy and inversion symmetry, resulting in zero $\sigma_{xy}^I$. When spins are canted, time-reversal symmetry is broken, so that the degeneracy is lifted and each band splits into two branches. Band crossing between branches from different original bands happens and contributes significantly to nonzero $\sigma_{xy}^I$ [Fig. S7]. In particular, the red and blue branches splitted from the same valance band cross each other near $\Gamma$ point. Meanwhile, the yellow band gradually gets closer to the crossing point with increasing canting angles, touching the blue band at the canting angle of $45°$. As a result, band hybridization and band inversion between these three branches occur, developing a small gap at $E = -84$ meV that causes the largest $\sigma_{xy}^I$. However, with the further rise of the yellow band, band inversion occurs again and the band crossing is recovered at $63.43°$. Eventually, at large fields, all spins are polarized along the c-axis. $\sigma_{xy}^I$ saturates to a fixed value that contributes to AHE $\sigma_{xy}^A$ in our analysis. The band structure and the corresponding $\sigma_{xy}^I$ at different canting angles are shown in Fig. S7, from which we can confirm that $\sigma_{xy}^I$ is indeed enhanced while the Fermi energy lies near the band crossing point or inside the band gap. Keeping the Fermi energy $E_F = -84$ meV fixed, the variation of the calculated $\sigma_{xy}^I$ as a function of canting angle is shown in Fig. 3(c). A peak at finite canting is addressed, consistent with the experimental observation in Fig. 2(e). The peak value is 95.7 S/cm, on the same scale as the experimental observation, but about two times larger, possibly attributed to the difference of eigenstate wavefunctions or repopulation in energy levels due to thermal fluctuations. Actually, the peak feature is persistent in a large window of the Fermi energy from $E_F = -40$ meV to $E_F = -100$ meV [Fig. S8]. We thus fully confirm the THE-like feature in EuCd$_2$As$_2$ is actually the intrinsic AHE.

The above calculations are based on the AFM ground states at T<T$_N$. The spins tilt from $0°$ to $90°$ with the increase of the magnetic field. $\sigma_{xy}^I$ reaches a maximum at the canting angle of $45°$, resulting in the pronounced enhancement of intrinsic AHE and thus the pronounced peak feature nonlinear to M. Since the influence of thermal fluctuation is small for T<T$_N$, the canting angle of all the spins is almost the same under a certain magnetic field. So $\sigma_{xy}^I$, approximating to $\rho_{xy}^{NA}/\rho_{xx}^2$, is almost temperature-independent [Fig. 2(c,f)]. A similar argument is also suitable for T>T$_N$. Although the magnetic ground state changes, the spins also tilt with the external magnetic field but the canting angle varies point by point due to the large thermal fluctuation. $\sigma_{xy}^I$ can be expressed as the weight average of that at different canting angles. So the peak values of $\sigma_{xy}^I$ [Fig. 2(f)] and $\rho_{xy}^{NA}/\rho_{xx}^2$ [Fig. 2(c)] become smaller at elevated temperatures due to more scattered distribution of the canting angles. The nonlinear AHE in EuCd$_2$As$_2$ at both T>T$_N$ and T<T$_N$ is thus well understood.

Finally, the momentum-space Berry curvature in EuCd$_2$As$_2$ is sensitive to not only the intensity but also the direction of the external magnetic field. The results are summarized in Figure 4 and the rotation geometry is depicted in the inset of Fig. 4(b). α is defined as the angle between the external magnetic field and [001] direction. α=0° and α=90° correspond to the out-of-plane and the in-plane magnetic fields, respectively. The longitudinal resistivity $\rho_{xx}(B)$ [Fig. 4(a)] increases monotonously with increasing α, showing an anisotropic behavior. But surprisingly, the peak value of the Hall resistivity $\rho_{xy}(B)$ [Fig. 4(b)] reaches the maximum at around α=30°, which is in contradiction with the conventional $\cos\alpha$ dependence of the anisotropic Hall resistivity. This provides another evidence of the important role that the spin canting plays in the band structure. Since the nonlinear AHE is nearly one order of magnitude larger than linear AHE at the peak position, here we simply show the $\rho_{xy}^{NA} + $ R$_s$M. The anomalous Hall angle and the approximated nonlinear AHE ratio at different α are shown in Fig. 4(d). $\rho_{xy}^{NA}$ and $\theta_{AH}$ can be greatly enhanced by tilting the magnetic field away from [001] direction, reaching maximum values of 430 μΩ·cm and 21% at α=30°, respectively, which is nearly 5 times larger than that at α=0°. Meanwhile, the nonlinear AHE always dominates the total Hall signals, with the approximate contribution ratio always larger than 80% from α=0° to α=90°. All the data shown in Fig. 4 were measured at T=2K. The angle-dependent $\rho_{xy}^{NA}$ and $\theta_{AH}$ at T=10K is shown in Supplementary Materials. They reach the maximum at α=30° as well, with a similar $\theta_{AH} = 20\%$ and even larger $\rho_{xy}^{NA} = 480$ μΩ·cm. This special angle-dependent feature possibly originates from a similar band structure evolution process with the spins tilt from in-plane AFM structure to forced FM structure along α=30° instead of [001] direction. But since this spin rotation process under a tilted external magnetic field is not well defined, it is not rigorous to calculate the corresponding band structure evolution here.

## III. CONCLUSION

In conclusion, we observed a giant nonlinear intrinsic AHE in EuCd$_2$As$_2$, which can be further enhanced by tilting the direction of the external magnetic field. Different from the common wisdom, this special Hall response originates from neither chiral spin textures nor the Weyl nodes, but is related to the specific spin-texture dependent band evolution. Our finding clearly unveils the sensitive dependence of band structure, *i.e.*, the momentum-space Berry curvature, on the real space spin orientations. It further indicates that this band structure evolution need to be taken into account not only in the study of AHE, but also other transport or topological properties of magnetic materials, thus provides a comprehensive understanding of the interplay between magnetic structures and electron properties.

## IV. METHOD

### A. Sample synthesis

Single crystals of EuCd$_2$As$_2$ were grown by Sn flux method. High-purity elements of Eu, Cd, As, and Sn were put in an alumina crucible at a molar ratio of 1:2:2:10 and sealed in a quartz tube under high vacuum. The tube was heated to 1173K, dwelt for 20 hours, and then slowly cooled to 773K at a rate of 2K/h. After that, the samples were separated from the Sn liquid in a centrifuge.

## B. Monte Carlo calculation

Spin interaction in the Monte Carlo simulations was modeled as

$$H = -\sum_{<i,j>} J\boldsymbol{m}_i \cdot \boldsymbol{m}_j - B\sum_i m_i^z$$

$$-\frac{16}{9}\sum_i K_1((\boldsymbol{m}_i \cdot \boldsymbol{u}^1)^2(\boldsymbol{m}_i \cdot \boldsymbol{u}_2)^2 + (\boldsymbol{m}_i \cdot \boldsymbol{u}_2)^2(\boldsymbol{m}_i \cdot \boldsymbol{u}_3)^2$$

$$+ (\boldsymbol{m}_i \cdot \boldsymbol{u}_3)^2(\boldsymbol{m}_i \cdot \boldsymbol{u}_1)^2)$$

$$-16\sum_i K_2((\boldsymbol{m}_i \cdot \boldsymbol{u}^1)^2(\boldsymbol{m}_i \cdot \boldsymbol{u}_2)^2(\boldsymbol{m}_i \cdot \boldsymbol{u}_3)^2),$$

where the first term is the nearest neighbor Heisenberg interaction, the second term is the Zeeman coupling, and the rest two terms are 6th order magnetocrystalline anisotropy, which is responsible for six preferred spin orientations in the plane. The parameters used in the simulations are $J_{xy} = 50k_B$, $J_z = -10k_B$, $K_1 = 2k_B$, $K_2 = 4k_B$, $B_z = 1.5k_B$, $u_1 = \left(-\frac{1}{2}, -\frac{\sqrt{3}}{2}, 0\right)$, $u_2 = (1,0,0)$, $u_3 = \left(-\frac{1}{2}, \frac{\sqrt{3}}{2}, 0\right)$. The system size is $256 \times 256 \times 2$. The periodical boundary condition was applied in both x- and y-directions, while an open boundary condition was used for the z- direction. All Monte Carlo simulations were carried out in our JuMag software, which is a GPU-accelerated package for spin dynamics and atomistic simulations [44]. The investigated systems were gradually cooled down from 500K to 0.1K, and 50000 Monte Carlo steps were performed for each temperature in order to sufficiently thermalize the system. At each sampled temperature, the topological charge distribution has been counted over 20000 samples. Two neighboring samples were separated by 500 Monte Carlo steps.

## C. Band structure calculations

The total energies and band structures were calculated from first-principles calculations within the framework of density functional theory using the projector augmented wave pseudopotential [36] as implemented in VASP [37,38]. The generalized gradient approximation of Perdew, Burke, and Ernzerhof [39] was used for the exchange-correlation energy and the Hubbard U method [40] with U = 6.0 eV and J = 1.0 eV was applied on the Eu(4f) orbitals. An energy cutoff 600 eV for the plane-wave expansion was used. Non-collinear magnetism calculations with spin-orbit coupling included were employed. A $1 \times 1 \times 2$ supercell with two Eu atoms was used and a Γ-centered $15 \times 15 \times 4$ $\boldsymbol{k}$-mesh was sampled. After we obtained the eigenstates and eigenvalues, a unitary transformation of Bloch waves was performed to construct the tight-binding Hamiltonian in a Wannier function basis by using the maximally localized Wannier functions method [41] implemented in the Wannier90 package [42]. WF-based Hamiltonian has the exactly same eigenvalues as those obtained by first-principles calculations from -0.5 ~ 0.5 eV to the Femi level. The intrinsic anomalous

Hall conductivity was calculated using the WF-based Hamiltonian based on Berry curvature [43].

## ACKNOWLEDGMENTS

F.X. was supported by the National Natural Science Foundation of China (11934005, 61322407, 11874116, 61674040), National Key Research and Development Program of China (Grant No. 2017YFA0303302 and 2018YFA0305601), the Science and Technology Commission of Shanghai (Grant No. 19511120500), the Shanghai Municipal Science and Technology Major Project (Grant No. 2019SHZDZX01), and the Program of Shanghai Academic/Technology Research Leader (Grant No. 20XD1400200). Work at UNH was supported by US Department of Energy, Office of Science, Basic Energy Sciences (Grant No. DE-SC0020221). Y.S. was supported by Chinese National Key Research and Development Program (No. 2017YFA0302901), the National Natural Science Foundation of China (U2032204, 12004416), the Strategic Priority Research Program (B) of the Chinese Academy of Sciences (No. XDB33000000). L.K. was supported by the National Natural Science Foundation of China (11974021). We thank Cheng Zhang and Xingtai Chen for insightful discussions.

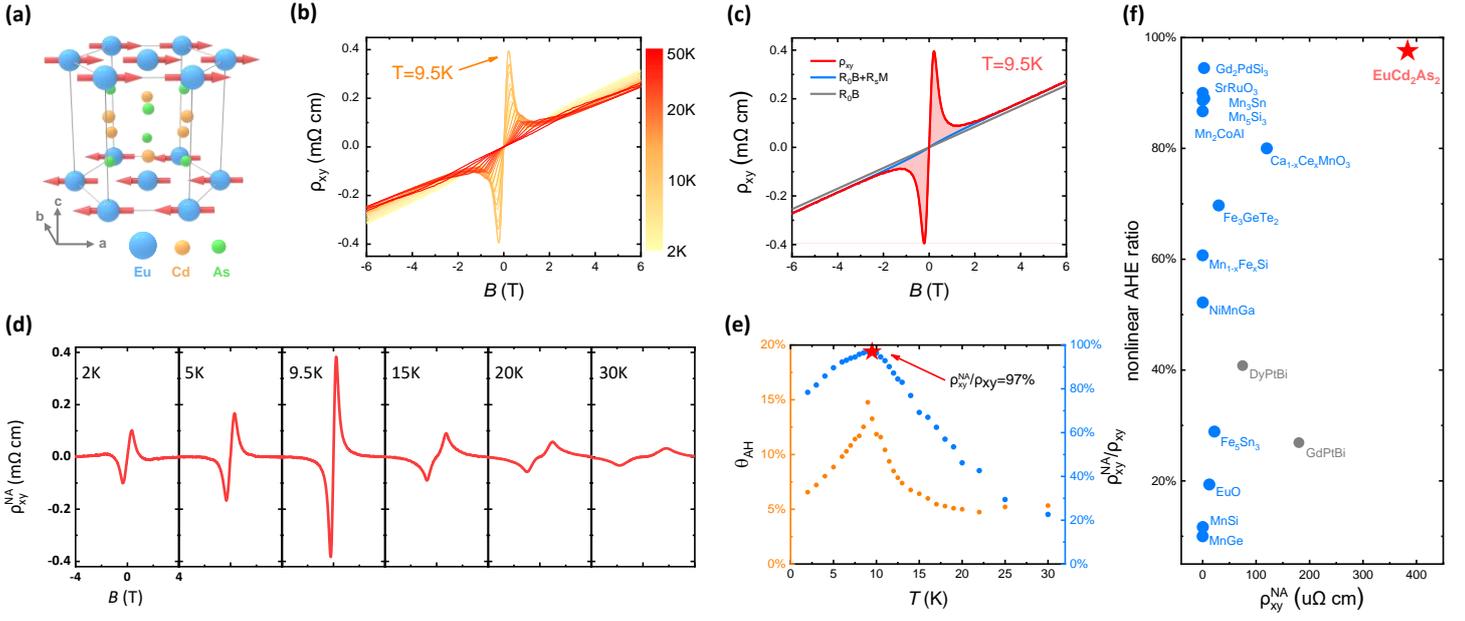

FIG. 1. Temperature-dependent nonlinear AHE in EuCd$_2$As$_2$. (a, b) Magnetic-field dependence of the Hall resistivity $\rho_{xy}$ at different temperatures. (c) Decomposition of the field-dependent Hall resistivity measured at T=9.5K. The red line is the original Hall resistivity. The blue line is the sum of the ordinary Hall effect and conventional AHE R$_0$B+R$_s$M. The red zone corresponds to AHE nonlinear to M. (d) Magnetic-field dependence of the nonlinear AHE resistivity $\rho_{xy}^{NA}$ at various temperatures from 2K to 30K. (e) Maximum anomalous Hall angle (yellow) and nonlinear AHE ratio (blue) at different temperatures. The nonlinear AHE contributes 97% of the total Hall signal at T=9.5K. (f) Nonlinear AHE ratio and $\rho_{xy}^{NA}$ of THE systems (blue) and nonlinear intrinsic AHE systems (grey). All data here were measured with B ∥ c and I ∥ a.

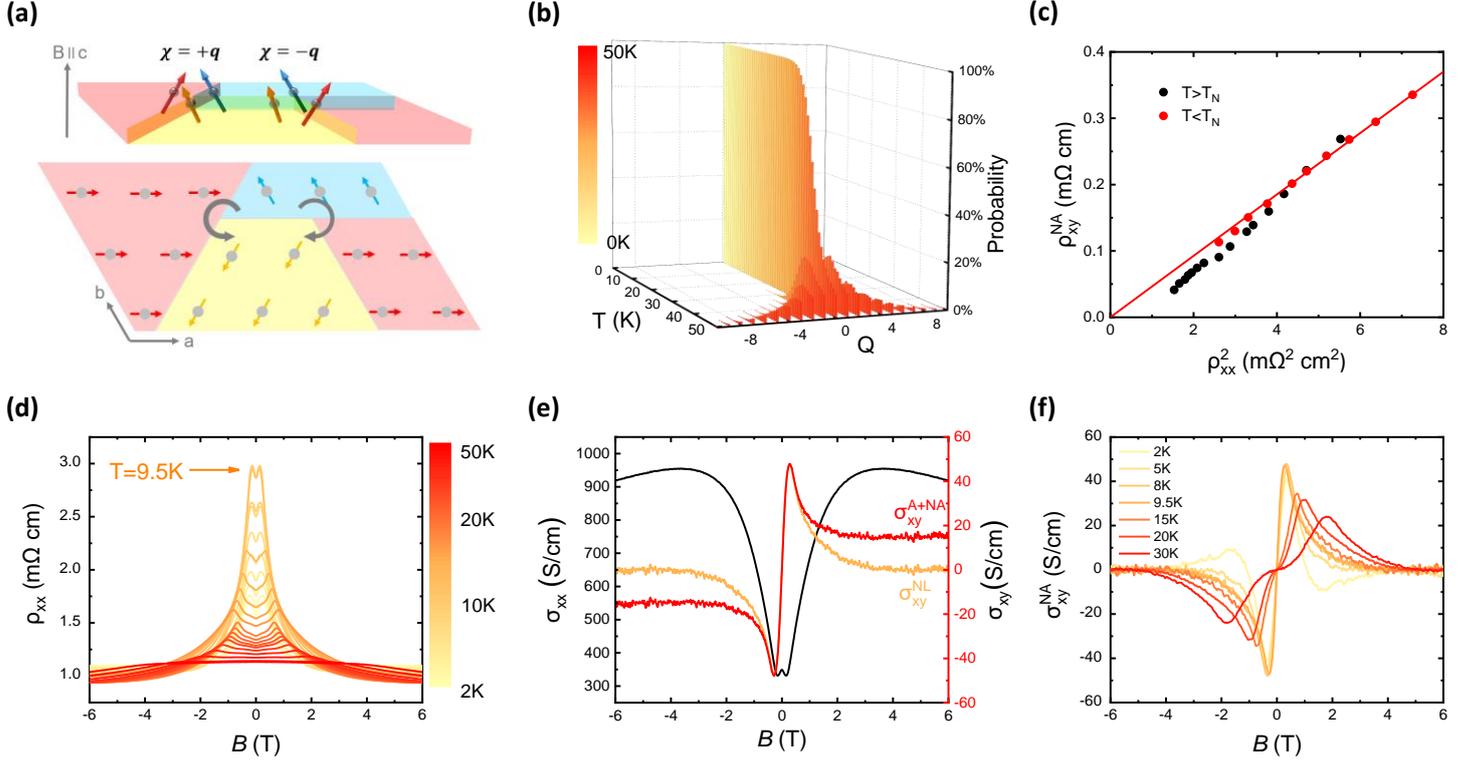

FIG. 2. Detailed analysis of nonlinear AHE in EuCd$_2$As$_2$. (a) Domain structure and the neutralization of overall scalar chirality in EuCd$_2$As$_2$. (b) Topological charges in EuCd$_2$As$_2$ at different temperatures. (c) Maximum $\rho_{xy}^{NL}$ as a function of $\rho_{xx}^2$. The red line is the linear fitting for T<T$_N$. (d) Magnetic-field-dependent longitudinal resistivity at different temperatures. (e) Magnetic-field-dependent longitudinal conductivity (black line), nonlinear anomalous Hall conductivity (yellow line), and non-ordinary Hall conductivity (red line) at T=9.5K. (f) Magnetic-field-dependent nonlinear anomalous Hall conductivity at different temperatures.

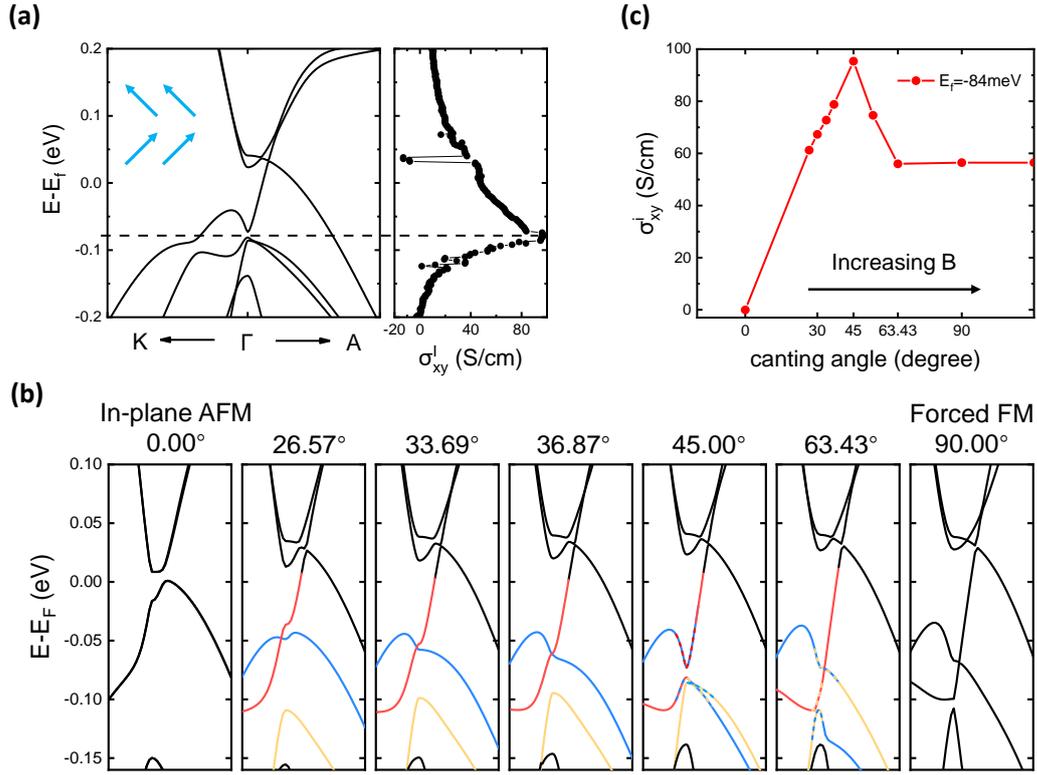

FIG. 3. Band structures and intrinsic Hall conductivity calculation results of EuCd$_2$As$_2$. (a) Calculated band structure along $K-\Gamma-A$ direction and the corresponding energy-dependent intrinsic Hall conductivities $\sigma_{xy}^I$ with the canting angle 45°. (b) Calculated band structures along $K-\Gamma-A$ direction with various canting angles. (c) Canting-dependent $|\sigma_{xy}^I|$ at $E_F = -84$ meV.

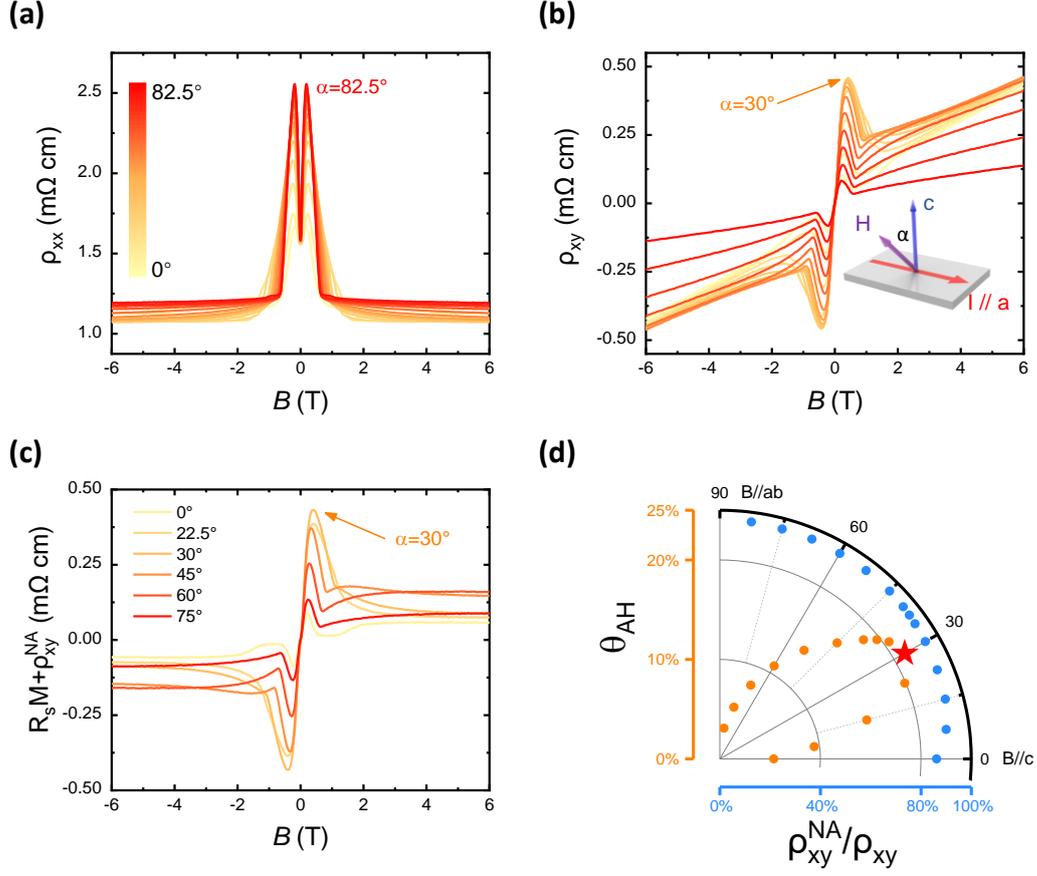

FIG. 4. Angle-dependent nonlinear AHE in EuCd$_2$As$_2$. (a) Magnetic-field dependence of the transverse resistance $\rho_{xx}$ at different angles. (b) Magnetic-field dependence of the Hall resistance $\rho_{xy}$ at different angles. (c) Total anomalous Hall resistivity at different angles. The maximum nonlinear AHE appears at around α=30°. (d) Maximum anomalous Hall angle (yellow) and approximate nonlinear AHE ratio (blue) at different angles. The anomalous Hall angle can reach 21% at α=30°. All data here were measured at T=2K.

15